\documentclass[sigconf]{acmart}

\acmConference[AFT'21]{Advances in Financial Technology}{September 26--28, 2021}{Arlington, VA, USA}

\acmYear{2021}\copyrightyear{2021}
\setcopyright{acmlicensed}
\acmConference[AFT '21]{3rd ACM Conference on Advances in Financial Technologies}{September 26--28, 2021}{Arlington, VA, USA}
\acmBooktitle{3rd ACM Conference on Advances in Financial Technologies (AFT '21), September 26--28, 2021, Arlington, VA, USA}
\acmPrice{15.00}
\acmDOI{10.1145/3479722.3480994}
\acmISBN{978-1-4503-9082-8/21/09}

\usepackage{graphicx}

\usepackage[caption=false,font=footnotesize]{subfig}
\usepackage[font=footnotesize,labelfont=bf]{caption}

\usepackage{makecell}

\usepackage{amsmath}
\usepackage{bbm}
\usepackage{stmaryrd}

\usepackage{array}
\usepackage{color}
\usepackage{multirow}
\usepackage{rotating}
\usepackage{makecell}
\usepackage{hhline}
\usepackage{lscape}
\usepackage{tabu}
\usepackage{tikz}
\usepackage{textcomp} 
\usetikzlibrary{shapes,backgrounds} 

\usepackage{colortbl}
\usepackage{wrapfig}

\usepackage{caption}

\usepackage{listings, xcolor}

\definecolor{verylightgray}{rgb}{.97,.97,.97}
\lstdefinelanguage{Solidity}{
keywords=[1]{anonymous, assembly, assert, balance, break, call, callcode, case, catch, class, constant, continue, contract, debugger, default, delegatecall, delete, do, else, event, export, external, false, finally, for, function, gas, if, implements, import, in, indexed, instanceof, interface, internal, is, length, library, log0, log1, log2, log3, log4, memory, modifier, new, payable, pragma, private, protected, public, pure, push, require, return, returns, revert, selfdestruct, send, storage, struct, suicide, super, switch, then, this, throw, transfer, true, try, typeof, using, value, view, while, with, addmod, ecrecover, keccak256, mulmod, ripemd160, sha256, sha3}, 
	keywordstyle=[1]\color{blue}\bfseries,
	keywords=[2]{Stages,States, address, bool, byte, bytes, bytes1, bytes2, bytes3, bytes4, bytes5, bytes6, bytes7, bytes8, bytes9, bytes10, bytes11, bytes12, bytes13, bytes14, bytes15, bytes16, bytes17, bytes18, bytes19, bytes20, bytes21, bytes22, bytes23, bytes24, bytes25, bytes26, bytes27, bytes28, bytes29, bytes30, bytes31, bytes32, enum, int, int8, int16, int24, int32, int40, int48, int56, int64, int72, int80, int88, int96, int104, int112, int120, int128, int136, int144, int152, int160, int168, int176, int184, int192, int200, int208, int216, int224, int232, int240, int248, int256, mapping, string, uint, uint8, uint16, uint24, uint32, uint40, uint48, uint56, uint64, uint72, uint80, uint88, uint96, uint104, uint112, uint120, uint128, uint136, uint144, uint152, uint160, uint168, uint176, uint184, uint192, uint200, uint208, uint216, uint224, uint232, uint240, uint248, uint256, var, void, ether, finney, szabo, wei, days, hours, minutes, seconds, weeks, years},	
	keywordstyle=[2]\color{teal}\bfseries,
	keywords=[3]{block, blockhash, coinbase, difficulty, gaslimit, number, timestamp, msg, data, gas, sender, sig, value, now, tx, gasprice, origin},	
	keywordstyle=[3]\color{violet}\bfseries,
	identifierstyle=\color{black},
	sensitive=false,
	comment=[l]{//},
	morecomment=[s]{/*}{*/},
	commentstyle=\color{gray}\ttfamily,
	stringstyle=\color{red}\ttfamily,
	morestring=[b]',
	morestring=[b]"
}

\usepackage{xspace}

\newcommand{\etc}{\textit{etc.}\xspace}
\newcommand{\ie}{\textit{i.e.,}\xspace}
\newcommand{\eg}{\textit{e.g.,}\xspace}
\newcommand{\cf}{\textit{cf.}\xspace}

\newcommand{\cmmnt}[1]{\ignorespaces} 

\usepackage{adjustbox}
\newcommand{\headrow}[1]{\multicolumn{1}{c}{\adjustbox{angle=45,lap=\width-0.5em}{#1}}}

\newcommand{\full}{$\bullet$}
\newcommand{\prt}{$\circ$}
\newcommand{\none}{$\times$}

\begin{document}

\title{SoK: Oracles from the Ground Truth to Market Manipulation}

\author{Shayan Eskandari}
\authornote{S. Eskandari and M. Salehi are equal first authors.}
\affiliation{\institution{Concordia University} \city{Montreal} \state{QC} \country{Canada}}
\affiliation{\institution{ConsenSys Diligence} \city{Brooklyn} \state{NY} \country{USA}}
\author{Mehdi Salehi}
\authornotemark[1]
\affiliation{\institution{Concordia University} \city{Montreal} \state{QC} \country{Canada}}
\author{Wanyun Catherine Gu}
\affiliation{\institution{Stanford University} \city{Stanford} \state{CA} \country{USA}}
\author{Jeremy Clark}
\affiliation{\institution{Concordia University} \city{Montreal} \state{QC} \country{Canada}}
\email{j.clark@concordia.ca}



\begin{abstract}

One fundamental limitation of blockchain-based smart contracts is that they execute in a closed environment. Thus, they only have access to data and functionality that is already on the blockchain, or is fed into the blockchain. Any interactions with the real world need to be mediated by a bridge service, which is called an oracle. As decentralized applications mature, oracles are playing an increasingly prominent role. With their evolution comes more attacks, necessitating greater attention to their trust model. In this systemization of knowledge paper (SoK), we dissect the design alternatives for oracles, showcase attacks, and discuss attack mitigation strategies.
    
\end{abstract}

\maketitle




\section{Introduction} \label{sec:intro}

With billions of dollars at stake, decentralized networks are prone to attacks. It is essential that the smart contracts, which govern how systems are run on these networks, are executed correctly. Public blockchains, like Ethereum, ensure the correct execution of smart contract code by taking the consensus of a large, open network of nodes operating the Ethereum software. For consensus to form, many nodes need to make decisions based on the exact same input data. Hypothetically, if a decision requires nodes to fetch data or use a service provider outside of the blockchain, there can be no guarantee that every node in a global network has the same access and view of this external source. For this reason, blockchains only execute on internal sources: data and code provided in a current transaction, or past data and code already stored on the blockchain.

Many potential decentralized applications seem very natural until the designer hits the `oracle problem' and realizes an interface to the external world is required. An oracle is a solution to this problem. It is a service that feeds off-chain data into on-chain storage. The trust model of oracles vary---some data comes with cryptographic certification while other data is assumed to be true based on trusting the oracle, or a set of oracles. Oracle-supplied data cannot easily be changed or removed once finalized on-chain, allowing disputes over data accuracy to be based on a public record. Leveraging this immutability is one approach to incentivizing oracles to post truthful information.

We aim to construct in this paper a systematization of knowledge (SoK) of implementation choices for oracles, facilitated by breaking down the operation of an oracle into a set of modules. For each module, we explore potential system vulnerabilities and discuss attack vectors. We also aim to categorize all the significant oracle proposals of different projects within a taxonomy we propose. The goal of this SoK is to help the reader better understand the system design for oracles across different use cases and implementations.

\section{Preliminaries}

Ethereum~\cite{wood2014ethereum} is a prominent public blockchain with the largest developer headcount. While oracles are applicable to any blockchain, we will adopt Ethereum as a concrete example of a blockchain for the purposes of explaining each concept in this paper. Ethereum is inspired by Bitcoin but adds a verbose language for programming \textit{smart contracts} that execute on the \texttt{Ethereum Virtual Machine (EVM)}. All transactions and executions are verified by a decentralized network of nodes. Solidity is the main high-level programming language used by developers for developing smart contracts and decentralized applications (DApps). Smart contracts are small code bases that live on a blockchain. In short, smart contracts can be seen as blackbox applications that get inputs from a user and follow the code flow to the output, which can update the state of the contract and trigger monetary transactions. 

\paragraph{The Oracle Problem.}  

Smart contracts cannot access external resources (\eg a website or an online database) to fetch data that resides outside of the blockchain (\eg a price quote of an asset). External data needs to be relayed to smart contracts with an oracle. An \emph{oracle} is a bridge or gateway that connects the off-chain real world knowledge and the on-chain blockchain network. The `oracle problem'~\cite{linkOracleProblem} describes the limitation with which the types of applications that can execute solely within a fully decentralized, adversarial environment like Ethereum. Generally speaking, a public blockchain environment is chosen to avoid dependencies on a single (or a small set) of trusted parties. One of the first oracle implementations used a smart contract in the form of a database (\ie mapping\footnote{A Solidity \texttt{mapping} is simply a key-value database stored on a smart contract.}) and was updated by a trusted entity known as the \texttt{owner}. More modern oracle updating methods use consensus protocol with multiple data feeds or polling techniques based on the `wisdom of the crowd'. The data reported by an oracle will always introduce a time lag from the data source and more complex polling methods generally imply longer latency.

\paragraph{Trusted Third Parties.} A natural question for smart contract developers to ask is: if you trust the oracle, why not just have it compute everything? There are a few answers to this question: (1) there may be benefits to minimizing the trust (\ie to just providing data instead of full execution), (2) there are widely trusted organizations and institutes---convincing one to operate an oracle service is a much lower technical ask than convincing one to operate a complete platform, and (3) if a data source becomes untrustworthy, it may require less effort to switch oracles than to redeploy the system. 

\paragraph{Methodology.} We found papers and other resources by examining the proceedings of top ranked security, cryptography, and blockchain venues; attending blockchain-focused community events; and leveraging our expertise and experience. Our inputs include academic papers, industry whitepapers, blog and social media posts, and talks at industry conferences on blockchain technology, Ethereum, and decentralized finance (DeFi). 

\paragraph{Oracle Use-Cases.} Oracles have been proposed for a wide variety of applications. Based on our reading, most of the use-cases fall into one of the main categories below.

\begin{itemize}

\item \textbf{Stablecoins}~\cite{clark2019sok,MSS20,PHP+19,gu2020empirical,MAKERDAOOracle} and \textbf{synthetic assets}~\cite{SCM21} require the exchange rate between the asset they are price-targeting and the price of an on-chain source of collateral. 
\item \textbf{Derivatives}~\cite{eskandari2017feasibility,biryukov2017findel,synthetix} and \textbf{prediction markets}~\cite{clark2014decentralizing,peterson2015augur} require external prices or event outcomes to settle on-chain contracts.
\item \textbf{Provenance systems}~\cite{RKYCC19,tian2016agri} require tracking information of real world assets like gold, diamonds, mechanical parts, and shipments.
\item \textbf{Identity}~\cite{KL17,maram2021candid} and other on-chain reputation systems require knowledge of governmental records to establish identities.
\item \textbf{Randomness}~\cite{chainlinkvrf} can only be produced deterministically on a blockchain. In order to use any non-deterministic random number, an external oracle is needed to feed the randomness into the smart contract. \textbf{Lotteries}~\cite{pooltogether} and \textbf{games}~\cite{etheroll} are examples. Additionally, cryptographic tools like verifiable random functions (VRF)~\cite{micali1999verifiable,goldbe-vrf-01} and verifiable delay functions (VDFs)~\cite{bunz2017proofs,crypto-2018-28858} can mitigate, respectively, any predictability or manipulability in the randomness.
\item \textbf{Decentralized exchanges} can use prices from an external oracles to set parameters. On-chain market makers~\cite{hertzog2017bancor} uses such prices to minimize the deviation from the external market prices and tailor the pricing function. Additionally, some use oracles to provide sufficient liquidity near the mid-market price for more efficient automated market making~\cite{dodopmm,cofixwhitepaper,cofixblog}.
\item \textbf{Dynamic non-fungible tokens (NFTs)}~\cite{chainlinknft} are crypto-collectables that can be minted, burned, or updated based on external data. For example, sports trading cards which depends on the real-time performance of a player.
\end{itemize}


\section{Related Work}

Given this paper is a systemization of knowledge (SoK), we will review work on oracles themselves throughout the paper. In this section, we only discuss other works with a similar goal of providing an overview of different approaches to oracle design, operation, and security. Al-Breiki et al. \cite{al2020trustworthy} present a trust-based categorization of oracle systems, as well as the type of interaction that the on-chain component of the oracle has with the off-chain components. 
Liu and Szalachowski \cite{liu2020first} focus on oracles in the decentralized finance (DeFi) ecosystem, presenting technical architectures and a measurement study on deviations between external market prices and on-chain data from commonly used price oracles.
Lo et al.~\cite{lo2020reliability} propose a framework for assessing the reliability of oracles and ranked them based on the failure probability rate.
Angeris and Chitra \cite{angeris2020improved} analyze the logic behind Automated Market Maker (AMM) projects (\eg Uniswap~\cite{adams2019uniswap} and Balancer~\cite{balancer}) and discuss how these projects could be used as price feeder oracles for other systems.
Williams and Peterson \cite{williams2019decentralized} map oracle systems into two groups---requesters and reporters--- and perform a game theoretical analysis of three defined scenarios between requesters and reporters.

By contrast, in our work, we inspect 17 different oracle systems\footnote{To our knowledge, a much larger set than other research on oracle systems.} and breakdown their design decisions and mechanism implementations (listed in Table~\ref{tab:classification}). We also discuss theoretical and possible attacks on the different building blocks of the oracle systems. Comparatively, we look at a broader types of oracles, including price oracles, binary outcome oracles, and oracle systems, for any type of data such as weather condition information. 


\section{Modular Work Flow} \label{overview_workflow}

For our main contribution, we deconstruct how an oracle operates into several modules that generally operate sequentially (but in some solutions, certain steps are skipped) and then we study each module one-by-one. An overview of the work flow is as follows:

\begin{description}
	\item \ref{ground_truth} \textbf{Ground Truth}: The goal of the oracle system is to relay the ground truth (\ie the real true data) to the requester of the data. 

	\item \ref{data_sources} \textbf{Data Sources}: Data Sources are entities that store or measure a representation of the ground truth. There are a diverse set of data sources: databases, hardware sensors, humans, other smart contracts, \etc

	\item \ref{data_feeders} \textbf{Data Feeders}: Data feeders report off-chain data sources to an on-chain oracle system. In order to incentivize truthful data reporting, an oracle system can introduce a mechanism to select data feeders from a collection of available data providers. The incentive mechanism can be collateral-based, such as staking, or reputation-based to find a reliable set of data feeders for each round of selection.
	
	\item \ref{data_feeder_selection} \textbf{Selection of Data Feeders}: The process of determining which data feeders should be used in an oracle system can be categorized into two main types: centralized and decentralized selection.

	\item \ref{aggregation} \textbf{Aggregation}: When data is submitted by multiple data feeders, the final representation of the data is an aggregation of each data feeder's input. The aggregation method can be random selection or algorithmic rule-based, such as using weighted average (the mean) or majority opinion. The design of the aggregation method is one of the most important aspects of an oracle system, as intentional manipulation or unintentional errors during the aggregation process can result in untruthful data reporting by the oracle system.
	
	\item \ref{dispute_phase} \textbf{Dispute Phase}: Some oracle designs allow for a dispute phase as a countermeasure to oracle manipulation. The dispute phase might correct submitted data or punish untruthful data feeders. The dispute phase might also introduce further latency.
	
	\end{description}
	
\begin{figure}[t!]
    \centering
    \includegraphics[width=0.35\textwidth]{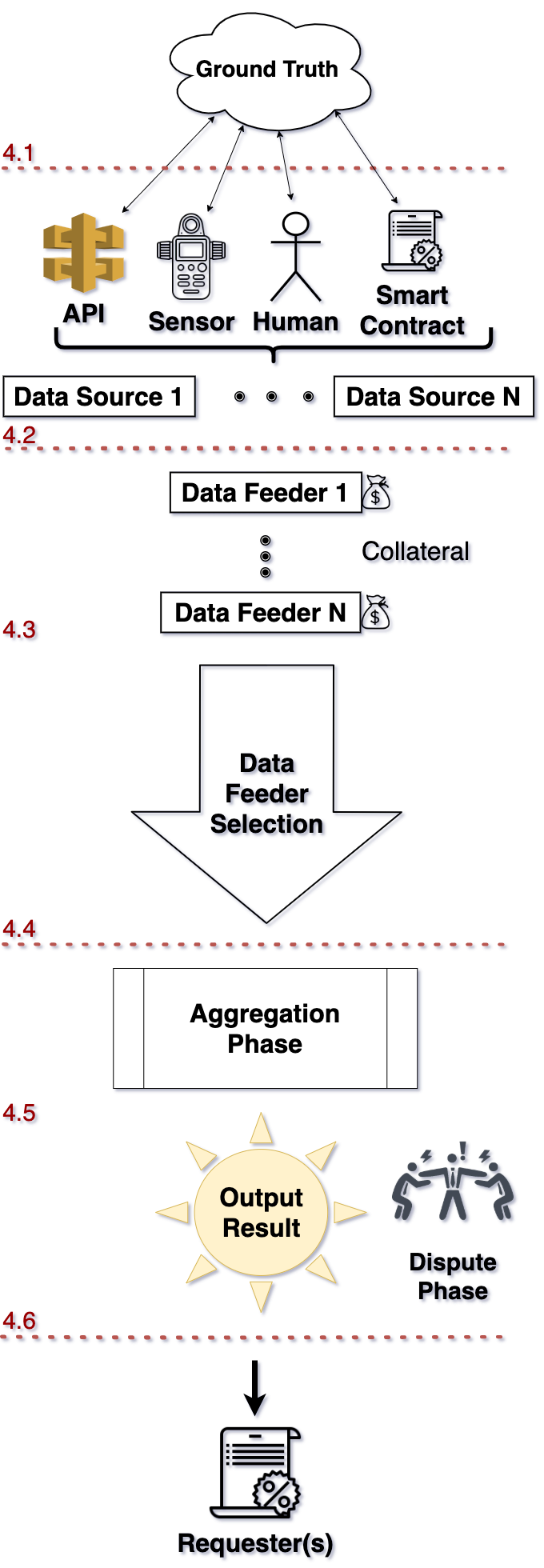}
    \caption{A visualization of our oracle workflow as described in the text.}
    \label{fig:dataflow}
\end{figure}

The steps above are visualized in Figure~\ref{fig:dataflow}. Next we dive deeper into the modular workflow by trying to further define each module. As appropriate, we also discuss feasible attacks on the modules and possible mitigation measures.


\subsection{Ground Truth}\label{ground_truth} 

While not a module itself, ground truth is the initial input to an oracle system. Oracle designers cannot solve basic philosophical questions like \textit{what is truth?} However it has to be understood (i) what the data actually represents and (ii) if it is reliable. Data is sometimes sensitive to small details. Consider a volatility statistic for a financial asset: basics like which volatility measure is being used over what precise time period are obvious, but smaller things like the tick size of the market generating the prices could be relevant~\cite{firat2002knowledge}. When data is aggregated from multiple sources, minor differences in what is being represented (called \emph{semantic heterogeneity}) can lead to deviations between values~\cite{madnick2006improving,worboys1991semantic,hakimpour2001resolving}.

While oracle systems will attempt to solve the issue of malicious participants who mis-report the ground truth, it does not address the fundamental question of whether the ground truth itself is reliable. Some philosophers argue truth is observed, and observations require a `web of beliefs' that is subject to error (for its consequences in security, see~\cite{HvO17}). Reliability is judged by the assumptions made about the data source, described next. 


\subsection{Data Sources}\label{data_sources}

Data Sources are defined here as passive entities that store and measure the representation of the ground truth. Common types of data sources include \texttt{databases}, \texttt{sensors}, \texttt{humans}, \texttt{smart contracts}, or a combination of them. Depending on how data sources gather and retrieve the ground truth, different attack types arrise. Using a hybrid of data sources (if possible) could reduce the reliability on a single point of input. We describe each common type and their security considerations.

\subsubsection{Humans.}\label{humansDatasource}
A human may provide the requested data, either by direct observation or by indirectly relaying data from another data source. Humans are prone to errors which is the main risk of this data source. Human errors include how the data is retrieved, how the data collector interprets the truth, and if data is relayed from a reliable source. Researchers have categorized human errors into the following three types (from least to most probable): very simple tasks, routine tasks, and complicated non-routine tasks~\cite{lo2020reliability}. An example for each category is, respectably, reading Bitcoin's exchange rate from an unverified source, inputting the data into the system, and configuring the oracle system.

Humans may also act maliciously and deliberately report wrong data when they perceive it will benefit them. As we will see in further modules, a robust oracle system will use incentives and disputes to promote truthful statements. 





\subsubsection{Sensors.} Sensors are electronic devices that collect raw data from the outside world and make it available to other devices. The data source may use more than one sensor to obtain the desired data. One example from traditional finance is the weather derivative, first introduced by the Chicago Mercantile Exchange (CME)~\cite{muller2000weather}. These instruments use weather data provided by trusted institutions, such as the National Climate Data Center,\footnote{\url{https://www.ncdc.noaa.gov/}} which collects weather data through a network of sensors.

Provenance is a highly cited application of blockchain, where products are tagged and traced through out the supply chain, including transportation, for management and/or certification~\cite{tian2016agri,mondal2019blockchain,zelbst2019impact}. The tags could be visual (barcodes) or electronic (RFID). A host of attacks on RFID have been proposed outside of blockchain oracles~\cite{alizadeh2012survey}. Blockchain technology does not solve some important trust issues: ensuring the proper tag is affixed to the proper product, each product has one tag, each tag is affixed to only one product, and tags cannot be transferred between products. This is called the \textit{stapling} problem~\cite{RKYCC19}. 

Sensors can produce noisy data or malfunction. The hardware of a sensor can also be modified when remote or physical access is unauthenticated (or weakly authenticated as many sensors are constrained devices). Probably the highest profile sensor attack (outside of blockchain) is Stuxnet~\cite{langner2011stuxnet}---malware that manipulated the vibration sensors, the valve control sensors, and the rotor speed sensors of Iran's nuclear centrifuges, causing the system to quietly fail~\cite{stuxnetattack}.

\subsubsection{Databases and application programming interfaces (APIs).} 

The most common mechanism used by software to fetch data is to use an API to obtain the data directly from a centralized database. A database is a set of tables that collect system events, while the API is an interface with the database. For example, a financial exchange keeps track of information in a database about every trade that has been executed. A data source that needs the daily traded volume of an asset could use the appropriate API of the exchange's database to extract the data from the related table in the database.

An active attacker can attack the system from two points. Modifying the data at rest in the database, or modifying the data in transit before and after the API call. 

\subsubsection{Smart Contracts.}

Smart contract could be used as a data source similar to a database. Decentralized finance (DeFi) applications on Ethereum include decentralized exchange services like Uniswap~\cite{adams2019uniswap}, or other oracles that operate on-chain. For instance API3 oracle~\cite{benligiraydecentralized} uses other on-chain oracles, called \textit{dAPIs}, as their data source. These oracles are whitelisted through voting by API3 token holders.

\textit{Automated Market Makers} (AMMs)~\cite{wang2020automated} are an on-chain alternative to centralized exchanges. Liquidity providers collateralize the contract with an equally valued volume of two types of cryptoassets. A mathematical rule governs how many assets of the one type are needed to purchase assets of the other. A well-known example of such mathematical rule is the \textit{Constant Function Market Makers} (CFMM) to calculate the exchange rates of tokens in a single trade~\cite{uniswapexplained}. The idea behind AMM was first raised by Hanson's logarithmic market scoring rule (LMSR) for prediction markets~\cite{hanson2003combinatorial}. A class of DeFi projects (\eg Uniswap~\cite{adams2019uniswap,adams2021uniswap} and Balancer~\cite{balancer}) uses CFMM to automate their market-making process. One of the utilizations of AMM is the ability to measure the price of an asset in a fully decentralized way, which addresses the \textit{pricing oracle problem}~\cite{angeris2020improved}.


One potential attack vector to the auto price discovery mechanism in an AMM is to manipulate prices provided by an algorithm, since the algorithmic rules used by an AMM is written in the smart contract and therefore how prices are quoted by the AMM can be calculated in advance. One real case example on \textit{bZx} is described in Section~\ref{oracle_smart_contract}. In addition to market manipulation, \textit{smart contract vulnerabilities}~\cite{atzei2017survey,chen2020survey} could possibly be used to influence the data coming from the Oracle, which we will discuss more in section~\ref{smart_contract}.



\subsection{Data Feeders}\label{data_feeders} 

Data feeders are entities who gather and report the data from a data source (Section~\ref{data_sources}) to the oracle system. 
A common configuration consists of an \textit{external feeder} which draws from off-chain data sources and deposit the data to an on-chain module. In case the data source is already on the blockchain, the data feeder step can be skipped.

It is not common to assume data feeders are fully honest, however a variety of threat models exist. Generally, this module will not attempt to determine if the data has been falsified (the later sections data selection (Section~\ref{data_feeder_selection}), data aggregation (Section~\ref{aggregation}) and dispute phase (Section~\ref{dispute_phase}) modules will deal with this issue); rather it will consist of tunnelling the data through the feeder with some useful security provisions. We discuss most important security provisions to achieve data integrity, confidentiality, and non-repudiation on any specific data.

\subsubsection{Source Authentication.}\label{source_authentication}
Data integrity can be enhanced by authenticating the source of the data and ensuring message integrity is preserved. It is sufficient to have the source sign the data, assuming the source's true signature verification (\ie public) key is known to the recipient of the data. This is most appropriate for sources like humans and sensors (although sensors may use a lightweight cryptographic alternative to expensive digital signatures~\cite{sallam2018survey}).

Databases, websites, and APIs typically support many cryptographic protocols, including the popular HTTPS (HTTP over SSL/TLS) which adds server authentication and message integrity to HTTP data~\cite{clark2013sok}. However HTTPS alone is typically not sufficient, as the message integrity it provides can only be verified by a client that connects to the server and engages in an interactive handshake protocol. This client cannot, for example, produce a transcript of what occurred and show it to a third party (\eg a smart contract on Ethereum) as proof that the message was not modified. To turn HTTPS data into signed data (or something similar), a trusted third party can vouch that the data is as received. TLS notary~\cite{tlsnotary} and DECO~\cite{zhang2019deco} offer solutions that attest for the authenticity of HTTPS data. Town Crier~\cite{zhang2016town} uses Trusted Execution Environments (TEE) like Intel SGX~\cite{costan2016intel} to push the trust assumption onto TEE technology and, ultimately, the chip manufacturer.

\subsubsection{Confidentiality.}


For many smart contracts that rely on oracles, the final data is made transparent (\eg prices, weather, event outcomes). In a few cases, oracles feed data that is private (\eg identities, supply chain information) and the contract enforces an access structure of which entities under which circumstances can access it~\cite{maram2021candid}.

Confidentiality might also be temporary.  Given the fact that information submitted to the mempool is public, there is a natural risk on the oracle system that a data feeder uses another data feeder's information to self-report to the system. This form of collusion between data feeders is called \textit{mirroring attack~\cite{ellis2017chainlink}} in computer security literature. The data feeders are willing to freeload another data feeder's response to minimize their cost of data provision. They will also be confident that their data will not be an outlier and be penalized. To mitigate the risk of mirroring attacks, the oracle designer should consider mechanisms that ensure the confidentiality of the data sent by the data feeders. A popular technique to achieve confidentiality is to use a commitment scheme~\cite{brassard1988minimum}. In a commitment scheme, each data feeder should send a commitment of the plain data as an encrypted message to the receiver. Later, the sender can reveal the original plain data and verify its authenticity using the commitment.

\subsubsection{Non-Repudiation.} 
A non-repudiation mechanism assures that a party cannot deny the sender's proposal after being submitted to the system. Oracle systems might rely on cryptographic signature schemes to eliminate the risk of in-transit corruption and to create irrefutable evidence of the data being provided by a source, for use in the dispute phase (Section~\ref{dispute_phase}) as needed.



\begin{table*}[t!]

    \renewcommand{\arraystretch}{1.3}
    
    \centering
    
    \begin{tabular}{llcccccc}
    
    \textit{Category} &
    \textit{Example} & 
    \headrow{No Trusted Third Party} & 
    \headrow{Low latency} &  
    \headrow{Resilient to Sybil Attacks} &
    \headrow{Resilient to Targetted DoS Attacks} & 
    \headrow{Incentives are Endogenous} & 
    
\\ \hline 
    
Centralized          & \texttt{Maker V1 Oracle}			&	&\full	&\full 	&	& 	\\
Voting  		& \texttt{Maker V2 Oracle}			&\full	& 	& 	&\full	& 	\\
Staking      		& \texttt{Chainlink}, \texttt{ASTRAEA} 	&\full	&\full	&\prt	&\full	&\full	\\ 
 	\hline
                                                                                       
    \end{tabular}
    
    \caption{Evaluation Framework on selection of data feeders. For details see Section~\ref{evaluation_framework}}
    \label{tab:data_feed_selection}
    \end{table*}

\subsection{Selection of Data Feeders}\label{data_feeder_selection} 

In order to ensure correct data is fed into the system, the design must select legitimate data feeders and weed out less qualified and malicious participants. In a non-adversarial environment, the design might aggregate all the incoming data without any selection, skipping this step.

The earliest designs for oracle systems, such as Oraclizeit~\cite{bernanioraclize} and PriceGeth~\cite{eskandari2017feasibility}, were designed using just one single data feeder; however, to improve data quality and the degree of decentralization, more complex oracle systems such as ChainLink~\cite{ellis2017chainlink} involves selecting qualified data feeders to aggregate an output that is expected to be more representative of the ground truth.

This process can be categorized into two main types: centralized and decentralized selection, with decentralized selection having multiple approaches through voting and staking. Centralized selection and decentralized selection through voting, create an allowlist of legitimate data feeders, in contrast to selecting based on the algorithmic criteria in decentralized selection through staking. 

\subsubsection{Centralized (Allowlist) Selection.} A centralized selection is a permissioned approach where a centralized entity selects a number of data feeders directly without the involvement of other participants in the network. A centralized selection is analogous to having an allowlist for authorized data feeds (\eg Maker Oracle V1~\cite{MAKERDAOOracle}). 
Compared to a decentralized approach, centralized selection is fast and direct. However the trust footprint on the central entity is large: it must solely select legitimate data feeders and also have high availability to update the allowlist as needed.

\subsubsection{Decentralized (Allowlist) Selection through Voting.} By decentralizing the selection process, the goal is to distribute the trust from a single entity to a collective decentralized governance.
Voting distributes trust and provides a degree of robustness against entities failing to participate, however it adds latency and introduces the threat that an actor can accumulate voting rights to sway the vote~\cite{makerdaoflashloanattack}, or even to do distrust and destroy the system (\eg Goldfinger attack~\cite{kroll2013economics}). 

For instance, in Maker V2 oracle~\cite{MAKERDAOOracle}, the selection of the data feeders is done through a decentralized governance process~\cite{gu2020empirical}. MKR\footnote{MakerDAO Governance Token} token holders vote on the number of authorized data feeders and who these data feeders can be~\cite{coinmonkMKRgovernance}.

Note that sometimes voting processes can provide the illusion of decentralization while not being much different than a centralized process in practice. To illustrate, consider a project with a governance token, in which most tokens are held by a few individuals where the project leaders advocate for their preferences and there is no established venue for dissenting opinions. If voters only inform themselves from one source of information, that source becomes a de facto centralized decision maker. 

\subsubsection{Decentralized Selection through Staking.} Like voting, staking attempts to utilize a token to align the incentives of the participants with the current functioning of the system. Mechanically, it works different: data feeders post collateral against the data they provide. In the dispute phase~\ref{dispute_phase}, any malicious data feeders will be punished by losing a portion or all of their collateral (called \textit{slashing}). Even without slashing, the collateral amount acts as a barrier to entry for participants and rate-limits participant.  

The stake can be both in token value and reputation of the data feeder. As an example, in Chainlink~\cite{ellis2017chainlink} protocol has a reputation contract that keeps track of the accuracy of data reporting of different feeders. The \texttt{ExplicitStaking} module in Chainlink 2.0 defines the number of Link tokens each oracle node must stake to become a data feeder, while the service agreement of the Chainlink oracle defines the circumstances in which a node's stake will be slashed~\cite{chainlinkExplicitStaking}. Put together, the incentives for selected data feeders to act honestly are avoiding reputational loss, avoiding loss of stake and penalty fees, and maintaining good standing for future income. In terms of selection, the data selection module forms a leaderboard, based on collateral and reputation, to select the highest ranked data feeders from all available feeders.

Another approach, introduced by ASTRAEA~\cite{adler2018astraea}, uses a combination of game theory and collateralization between different actors in the system (Voters and Certifiers) to achieve equilibrium on what the final data should be. 

A staking-based selection module avoids a central trusted third party, but it can add latency for adding/remove data feeds and other adjustments. It is also open to sybil attacks by design, while working to ensure these attacks have a significant cost for the adversary. 

One challenge for designing a staking mechanism is setting a high enough punishment (slashing) mechanism to thwart malicious actions. Projects like UMA~\cite{umawhitepaper}, another smart contract oracle design, dynamically adjust staked collateral needed for each round to ensure that \textit{Cost of Corruption} (CoC) is higher than the projected \textit{Profit from Corruption} (PfC). Profit from Corruption is defined by the data requester, in which UMA contracts require higher collateral to finalize the data from the data feeders. It is also important that participants are incentivized to file correct disputes---ones that will ultimately lead to identifying misbehaviour. If disputes are filed on-chain, the disputer will have to pay gas costs that need to be ultimately reimbursed by the resolution process. 



Decentralized selection is done by the holders of some scarce token, typically a governance token specific to the oracle service. The simplest decentralized mechanism to hold a vote amongst token holders, who are indirectly incentivized (we call this an \textit{exogenous incentive}) cast informed votes since they hold a token tied to the success of the system (\eg TruthCoin~\cite{sztorc2015truthcoin}). In a staking system, token holders are directly incentivized (a \textit{endogenous incentive}) to vote `correctly' (this remains to be defined but assume for now it means they vote in a way that will not be disputed) by posting some amount of their tokens as a fidelity bond. Stakers stand to be rewarded with new tokens and/or penalized (collateral slashed) depending on the performance of the data feeders they vote for.


Additionally a protocol could introduce a random selection within the data feeders to decrease the chance of sybil attacks. As an example Band Protocol~\cite{bandwhitepaper}, choses a random validator from top 100 staked participants for their oracle system. 

Another approach used by Tellor oracle~\cite{tellorWhitepaper} is a simple Proof of Work (PoW) algorithm for each round of data. The first 5 miners to submit their desired data alongside the solution to the mining puzzle are selected as the data feeders of the round. The selection is based on the hash power of each data feeder and randomness nature of proof of work consensus.

\subsubsection{Evaluation Framework on the selection of data feeders}\label{evaluation_framework}
To compare designs for data feeder selection, we provide an evaluation framework. The definition of each evaluation criteria (\ie column of the table) follows, specifying what it means to receive a full dot (\full), partial dot (\prt) or to not receive a dot. 

\emph{No Trusted Third Party.} \textit{A selection process that is distributed or decentralized among several equally-powerful entities earns a full dot (\full). A process that relies on a single entity for critical functions is not awarded a dot.}

The voting and staking processes are decentralized amongst multiple token holders (\full). As the name implies, the centralized process uses a trusted third party (no dot).

\emph{Low Latency.} \textit{A selection process that can move from proposal to finality within a single transaction is awarded a full dot (\full). A process that requires multiple rounds of communication or communication among several entities is not awarded a dot.}

The centralized process can make selection decisions unilaterally (\full). The voting process involves a round of communication with all of the participants (no dot). The staking process draws feeders unilaterally from an established leaderboard (\full). 

\emph{Resilient to Sybil Attacks.} \textit{A selection process that only allows unique feeders to participate is awarded a full dot (\full). The evaluation does not consider what specific method is used to determine entities are unique but assumes it works reasonably well (not strictly infallible). A process that is open to multiple fake feeders controlled by the same adversary is awarded a partial dot (\prt) if each additional feeder created by the adversary has a material financial cost. If there is no material cost to creating additional fake feeders, the process receives no dot.}

The centralized process manages an allowlist based on real world reputations. We assume this reasonably prevents sybils (\full). The staking process admits sybils but deters them by requiring staked tokens for each, which is costly (\prt). The voting process does not deter sybils from entering the election but relies instead on the voting process to not select them (no dot).   

\emph{Resilient to Targeted Denial of Service Attacks.} \textit{A selection process that only halts when multiple entities to go offline or fail is awarded a full dot (\full). If critical functionalities cannot be performed with the failure of a single entity, but the basic selection process can proceed, it is awarded a partial dot (\prt). If the process can be fully halted by the failure of a single entity, it is awarded no dot.}

The voting and staking processes can proceed until enough honest participants fail that a dishonest majority remains (\full). By contrast, a failure with the central entity in a centralized process can prevent critical functionalities, like updating the allowlist (\prt).

\emph{Incentives are Endogenous.} \textit{Every selection process should have the ability to remove untruthful feeders. Some selection processes might go beyond this and incentivize feeders to provide truthful information. Processes are awarded a full dot (\full) if the awards/punishments can be realized by the selection process itself. If the selection process relies only on external incentives (\eg damage to reputation), it is awarded no dot. The evaluation does not consider how information is determined to be truthful or not. Endogenous means the design is simpler but does not imply it is more secure (\cf \cite{FoBo19}).}

The staking process requires feeders to post collateral that can be taken (\ie slashed) for malicious behavior (\full). Centralized and voting processes do not use internal incentives (no dot).


\subsection{Aggregation of Data Feeds}\label{aggregation} 

Aggregation is the process of synthesizing the selected data feeds into one single output. The quality of the output depends on the data feed selection (see Section~\ref{data_feeder_selection}) and the aggregation process used. To highlight the importance of designing an aggregation method correctly, consider the case of Synthetix, a trading platform~\cite{synthetix} that used the \textit{average} (or \textit{mean}) of two data feeders as their aggregation method. An attacker leveraged this to manipulate one of the two feeders by inflating the real price by 1000x. Mean aggregation is highly sensitive to outlier data and the attack resulted in Synthetix's loss of several million dollars~\cite{synthetixIncident}.

\subsubsection{Statistical Measures.}

The three core statistics for aggregation are mean, median and mode. Many oracle systems use the median as the aggregated output, by selecting the middle entry of a list of ordinal data inputs. Unlike the mean, the median is not skewed by  outliers, although it assumes the inputs have an appropriate statistical distribution where the median is a representative statistic for the underlying ground-truth value. For example, if we believe data from the feeders is normally distributed with possible outliers, the median is appropriate. However if we believe it is bi-modally distributed, then discretizing and computing the mode (most common value) of the data is more appropriate. The mode is useful for non-numeric data (and nominal numbers). An approximation to the mode is picking a data input at random, however access to randomness from a smart contract is a well-documented challenge~\cite{chatterjee2019probabilistic,bunz2017proofs,chainlinkvrf}. Oracle projects like Chainlink do not prescribe a fixed aggregation method and let the data requesters select one.

To improve the quality of simple statistics such as the median and the mode, weights can be applied in the calculation. For instance, to mitigate manipulation of price data, one can choose to use \textit{time-weighted average price} (TWAP)~\cite{uniswaporacle}, or liquidity volume, or both~\cite{adams2021uniswap}. Typically, the liquidity and trading volume of a market correlates with the quality of the price data. To illustrate, Uniswap V2 uses TWAP over several blocks (\eg mean price in the last 10 blocks) to reduce the possibility of market manipulation in a single block (\eg via flash loans~\cite{qin2020attacking}). In Uniswap V3, TWAP is optimized for more detailed queries including the liquidity volume and allowing users to compute the geometric mean TWAP~\cite{adams2021uniswap}.

\subsubsection{Stale Data.}

Some use cases require frequent updates to data, such as weather data and asset prices. Stale data can be seen as valid data and pass the selection criteria, but it will reduce the aggregated data quality. Projects like Chainlink rank feeders based on historic timeliness. A naive approach ignores this issue and always uses the last submitted data of a data feeder even if the data feeder has not updated its price for some specific period. This approach is problematic if the underlying data is expected to change frequently. An example occurred on Black Thursday 2020~\cite{blackthursdayMaker} to MakerDao when Maker's data feeders could not update their feeds because of very high network congestion. After a significant delay in time, feeds were updated. The price had shifted by a large amount and the reported data jumped, leading to sudden, massive liquidations that were not adequately auctioned off. 



\subsection{Dispute Phase}\label{dispute_phase} 



The dispute phase is used to safeguard the quality of the final output and give the stakeholders a chance to mitigate inclusion of wrong data. Dispute resolution can be an independent module after the aggregation phase or it can be implemented at any other oracle module (\eg at the end of every aggregation~\ref{aggregation} or data feeder selection~\ref{data_feeder_selection}). Most oracle systems do dispute resolution internally, but market specialization has produced firms that provide outsourced dispute resolution as service (\eg Kleros~\cite{kleros}). To systemize the landscape, we first distinguish between systems that aim to detect (and remove) bad data providers and systems that vet the data itself. We then iterate how data is determined to be valid or invalid for the purposes of a dispute. Finally, we illustrate the consequences of a successful dispute: what happens to the disputed data and what happens to its provider.

\subsubsection{Provider-level and Data-level Vetting}\label{provideranddatavetting}

Dispute resolution can be \textit{provider-oriented} or \text{data-oriented}. Under a \textit{provider-oriented} regime, the focus is on selecting honest data providers and using disputes to remove data providers from serving as oracles in the future. In the optimistic case that providers are honest, oracle data is available immediately, however if an honest provider is corrupted, it will have a window of opportunity to provide malicious data before being excluded. One illustration of a provider-oriented system is operating a centralized allowlist of data providers (\eg MakerDAO v2) where providers can be removed. Chainlink~\cite{ellis2017chainlink}  strives to decentralize this functionality, where a reputation-based leaderboard replaces the allowlist. 

In a \text{data-oriented} regime, the focus is vetting the data itself. This can result in a slower system as oracle data is staged for a dispute period before it is finalized, however it can also correct false data (not merely remove the corrupted data feeder from future submissions). One illustration of a data-oriented system is Tellor~\cite{tellorWhitepaper,tellordispute}, where data is staged for 24 hours before finalization. If it is disputed, a period of up to 7 days is implemented to resolve the dispute. It is also possible that a system allows the resolution itself to be further disputed with one or more additional rounds. In Augur~\cite{peterson2015augur} for instance, the dispute step may happen in one round (takes maximum 1 day) or may contain other rounds of disputes that can last more than 7 days. 


\subsubsection{Determining the Truth}

In the optimistic case, an oracle system will feed and finalize truthful data, while disputes enable recourse for incorrect data. However disputes also introduce the possibility of two types of errors.

\begin{center}
\begin{tabular}{l|l|l|}
& No Disputes & Disputed \\ \hline
Data is correct & Correct & False Positive \\
Data is incorrect & False Negative & Correct \\
\end{tabular}
\end{center}

Dispute resolution in oracle systems focus on false positives. Incentivizing the discovery of false positives is present in some staking-based systems, however false negatives are not otherwise dealt with. In order to resolve a false positive, correct data must be used as a reference but, of course, if correct data is available as a reference, then it could replace the entire oracle system. That leaves two reasons for why an oracle system might still exist: (a) the reference for correct data is too expensive to consult on a regular basis, or (b) there is no reference for correct data and it must be approximated. 

If feeders are placed on an allowlist by a trusted party, disputes could be filed with the trusted party and manually verified. As far as we know, this is the only example of (a), although (a) is the basis for other blockchain-based dispute resolution protocols like optimistic roll-ups~\cite{KGCWF18}. The rest of the truth discovery mechanisms are based on (b) approximating the truth. 

A \textit{statistical approach} is selecting, from a set of values proposed by different feeders, the median of the values (\eg appropriately distributed continuous numerical data) or the mode (\eg non-continuous or non-numerical data). It is possible to augment this approach by having feeders \textit{stake} collateral in some cryptocurrency (\eg a governance token for the oracle project), and this collateral is taken (\textit{slashed}) from the feeder if their data deviates from the median by some threshold. If the amount slashed is payed, in part or in full, to the entity that filed and/or supported a dispute on the data, this incentivizes feeders to help reduce false negative errors in addition to false positives. One challenge is setting an acceptable threshold for slashing. A large threshold tolerates moderately incorrect data without punishment, while a small threshold could punish data feeders that are generally honest but faulty, slow, or reporting on highly volatile data. 

If a governance token exists for the oracle project, a related approach is to introduce \textit{voting} on disputed data by any token holder, and not limit the decision to just the feeders. In Augur~\cite{peterson2015augur} and ASTRAEA~\cite{adler2018astraea}, disputers vote to change the tentative outcome because they believe that outcome is false. Voting occurs over a window of time which extends the time to resolve disputes. By comparison, statistical mechanisms can be applied automatically and nearly instantly after the data is aggregated. However voting incorporates human judgement which might produce better outcomes in nuanced situations.

One final truth discovery mechanism is \textit{arbitrage} which is applicable in the narrow category of exchange rates between two on-chain tokens. This can be illustrated by the NEST oracle~\cite{nestwhitepaper} where data feeders assert the correct exchange rate between two tokens by offering a minimum amount of both tokens at this rate (\eg 10 ETH and 39,000 USDT for a rate of ETH/USDT = 3900). If the rate is incorrect, other participants will be given an arbitrage opportunity to buy/sell ETH at this rate, an action that can correct the price. This is very similar to drawing a price from an on-chain exchange, like Uniswap, and suffers from he same issue: an adversary can manipulate the oracle price by spending money. It is secure when the \textit{Cost of Corruption} (CoC) is greater than the \textit{Profit from Corruption} (PfC), however PfC can never be adequately accounted for because profit can come from extraneous (extra-Ethereum) factors~\cite{FoBo19}. The UMA~\cite{lambur2019data} oracle system has data feeders provide their own PfC estimates for the data they provide. 

\subsubsection{Consequences for Incorrect Data}

We now consider the consequences for disputed data that has been determined to be incorrect. In provider-oriented dispute resolution, incorrect data has consequences for the data feeder (see next subsection) but not the data itself. By the time the dispute is resolved, it is \textit{too late} to change the data itself. 

In data-oriented dispute resolution, data that has been deemed incorrect can either be \textit{reverted} or \textit{corrected}. Reversion means the outcome result will be annulled and the system should start from scratch to obtain new data, while corrected data will reflect a new undisputed value. The difference between the two is essentially in the complexity of the dispute resolution system. For reversion, a collective decision is taken to accept or reject data --- a binary option that is known in advanced. By contrast, correcting data requires new data to be proposed and then a collective decision to be made on all the proposals which is more complex but does not avoids rerunning the oracle workflow.

These differences also impact \textit{finality}: when should oracle data be considered usable? Dispute periods, re-running the workflow, and allowing resolved disputes to be further disputed can all introduce delays. To illustrate, consider  Augur~\cite{peterson2015augur} which implements a prediction market on binary events. Any observer with an objection to a tentative outcome can start a dispute round by staking REP (Augur's native token) on the opposite outcome. Dispute windows are 24 hours and then extended to 7 days for disputes on disputes. If the total staked amount exceeds 2.5\% of all REP tokens, the market enters a 60-day settlement phase called a fork window when all REP holders are obliged to stake on the final outcome. 

\subsubsection{Consequences for Data Feeder}

If data has been deemed incorrect through disputes or rejected for being an outlier, the feeder who provided the data might face consequences like being banned, slashed, or suffering reputational loss. It is also possible that there is \textit{no consequence} for the feeder other than the data being discarded. For example, in a sensor network, results from faulty sensors could have their data filtered out but continue to contribute data in expectation that they will be repaired in the future. 

In oracle designs based on allowlists, a feeder could be \textit{banned} or temporarily suspended for providing incorrect data. For dispute resolution based on staking, a feed could suffer \textit{economic loss} by having their stake taken from them. It is important to reiterate that this economic loss does not necessarily outweigh the utility of attempting to corrupt oracle data. The profit from corruption depends on where the data is being used, which could be within larger system than the blockchain itself~\cite{FoBo19}. Finally, a feeder might suffer \textit{reputational loss} for providing incorrect data. One can imagine this would be the case if, for example, the Associated Press misreported the outcome of the 2020 US Presidential election after announcing that it would serve as an oracle for this event on Ethereum.

Another illustration of these options is Chainlink, which maintains a decentralized analogue to a leaderboard where feeders are ranked according to the amount of LINK (Chainlink's token) they stake, as well as their past behavior in providing data that is timely and found to be correct. Data feeders with the outlier data will be punished by losing their collateralized LINK tokens and reducing their reputation score on the reputation registry. The lost of tokens is a direct cost, while the loss of reputation could impact their future revenue.


\subsection{Classification of Current Oracle Projects} \label{oracle_categories}


\begin{table*}[t!]
\centering

	\begin{tabular}{lllllllll}

&
\headrow{Data Source} &

\headrow{Selection Mechanism} & 
\headrow{Staking} &

\headrow{Aggregation Mechanism} &

\headrow{\underline{P}rovider/\underline{D}ata Vetting} &
\headrow{Determining the Truth} &
\headrow{Consequences (\underline{S}lash, \underline{B}an, \underline{L}oss)}  \\

\hline

		\multicolumn{1}{c|}{\textit{Oracle} } &   \multicolumn{1}{c|}{} &   \multicolumn{2}{c|}{\makecell{Data Feeder}} &  \multicolumn{1}{c|}{} &  \multicolumn{3}{c}{Dispute} &  \multicolumn{1}{c}{} \\
	
	\hline

	\makecell{ChainLink~\cite{ellis2017chainlink}}	& \multicolumn{1}{|c|}{API}  & \multicolumn{1}{c|}{\makecell{Reputation, \\ Staking}} &  \multicolumn{1}{c|}{\full} & \multicolumn{1}{c|}{\makecell{Statistical \\ Measure}} & \multicolumn{1}{c|}{P} & \multicolumn{1}{c|}{\makecell{Statistical \\ Measure}} & \multicolumn{1}{c}{S} &  \multicolumn{1}{c}{}\\

	\hline

	\makecell{UMA~\cite{umawhitepaper}}	& \multicolumn{1}{|c|}{Human, API}  & \multicolumn{1}{c|}{FCFS$^{\dagger}$} &  \multicolumn{1}{c|}{\full} & \multicolumn{1}{c|}{\none} & \multicolumn{1}{c|}{D} & \multicolumn{1}{c|}{Staking} & \multicolumn{1}{c}{S} &  \multicolumn{1}{c}{}\\

	\hline

	\makecell{Augur~\cite{peterson2015augur}}	& \multicolumn{1}{|c|}{Human}  & \multicolumn{1}{c|}{\makecell{Single \\ Source$^{\star}$}} &  \multicolumn{1}{c|}{\full} & \multicolumn{1}{c|}{\none} & \multicolumn{1}{c|}{D} & \multicolumn{1}{c|}{Voting} & \multicolumn{1}{c}{S} &  \multicolumn{1}{c}{}\\

	\hline

	\makecell{Uniswap~\cite{uniswaporacle}}	& \multicolumn{1}{|c|}{Smart Contract}  & \multicolumn{1}{c|}{\none} &  \multicolumn{1}{c|}{\none} & \multicolumn{1}{c|}{TWAP} & \multicolumn{1}{c|}{\none} & \multicolumn{1}{c|}{\none} & \multicolumn{1}{c}{\none} &  \multicolumn{1}{c}{}\\

	\hline

	\makecell{MakerDAO V1~\cite{MAKERDAOOracle}}	& \multicolumn{1}{|c|}{Human, API}  & \multicolumn{1}{c|}{\makecell{Centralized \\ Allowlist}} &  \multicolumn{1}{c|}{\none} & \multicolumn{1}{c|}{Median} & \multicolumn{1}{c|}{\none} & \multicolumn{1}{c|}{\none} & \multicolumn{1}{c}{\none} &  \multicolumn{1}{c}{}\\

	\hline

	\makecell{MakerDAO V2~\cite{MAKERDAOOracle}}	& \multicolumn{1}{|c|}{Human, API}  & \multicolumn{1}{c|}{\makecell{Decentralized \\ Allowlist}} &  \multicolumn{1}{c|}{\none} & \multicolumn{1}{c|}{Median} & \multicolumn{1}{c|}{P} & \multicolumn{1}{c|}{Voting} & \multicolumn{1}{c}{B} &  \multicolumn{1}{c}{}\\

	\hline

	\makecell{NEST~\cite{nestwhitepaper}}	& \multicolumn{1}{|c|}{Human}  & \multicolumn{1}{c|}{\none} &  \multicolumn{1}{c|}{\full} & \multicolumn{1}{c|}{\none$^{\star\star}$} & \multicolumn{1}{c|}{D} & \multicolumn{1}{c|}{Arbitrage} & \multicolumn{1}{c}{L} &  \multicolumn{1}{c}{}\\

	\hline

	\makecell{Band protocol~\cite{bandwhitepaper}}	& \multicolumn{1}{|c|}{API}  & \multicolumn{1}{c|}{\makecell{Random \\ Selection}} &  \multicolumn{1}{c|}{\full} & \multicolumn{1}{c|}{\makecell{Statistical \\ Measure}} & \multicolumn{1}{c|}{P} & \multicolumn{1}{c|}{Staking} & \multicolumn{1}{c}{S} &  \multicolumn{1}{c}{} \\

	\hline

	\makecell{Tellor ~\cite{tellordispute}}	& \multicolumn{1}{|c|}{Human, API}  & \multicolumn{1}{c|}{PoW} &  \multicolumn{1}{c|}{\full} & \multicolumn{1}{c|}{Median} & \multicolumn{1}{c|}{P} & \multicolumn{1}{c|}{Staking} & \multicolumn{1}{c}{\makecell{S \\ B}} &  \multicolumn{1}{c}{} \\

	\hline

	\makecell{ASTRAEA~\cite{adler2018astraea} \\ TruthCoin ~\cite{sztorc2015truthcoin}}	& \multicolumn{1}{|c|}{Human}  & \multicolumn{1}{c|}{Staking} &  \multicolumn{1}{c|}{\full} & \multicolumn{1}{c|}{Mode} & \multicolumn{1}{c|}{D} & \multicolumn{1}{c|}{Voting} & \multicolumn{1}{c}{S} &  \multicolumn{1}{c}{} \\

	\hline

	\makecell{Provable~\cite{bernanioraclize} \\ PriceGeth ~\cite{eskandari2017feasibility}}	& \multicolumn{1}{|c|}{API}  &   \multicolumn{1}{c|}{\none} & \multicolumn{1}{c|}{\none} & \multicolumn{1}{c|}{\none} & \multicolumn{1}{c|}{\none} & \multicolumn{1}{c|}{\none} & \multicolumn{1}{c}{\none} &  \multicolumn{1}{c}{} \\

	\hline

	\makecell{DIA Oracle ~\cite{DIAOracle}}	& \multicolumn{1}{|c|}{\makecell{API, \\ Smart Contract}}  &   \multicolumn{1}{c|}{\none} & \multicolumn{1}{c|}{\none} & \multicolumn{1}{c|}{\none} & \multicolumn{1}{c|}{D} & \multicolumn{1}{c|}{Staking} & \multicolumn{1}{c}{B} &  \multicolumn{1}{c}{} \\

	\hline

	\makecell{DECO ~\cite{zhang2019deco} \\ TownCrier ~\cite{zhang2016town}}	& \multicolumn{1}{|c|}{\makecell{HTTPS}}  &   \multicolumn{1}{c|}{\none} & \multicolumn{1}{c|}{\none} & \multicolumn{1}{c|}{\none} & \multicolumn{1}{c|}{\none} & \multicolumn{1}{c|}{\none} & \multicolumn{1}{c}{\none} &  \multicolumn{1}{c}{} \\

	\hline

	\makecell{API3~\cite{benligiraydecentralized} $\char`\\$w Kleros~\cite{kleros}}	& \multicolumn{1}{|c|}{Oracles}  &   \multicolumn{1}{c|}{\makecell{Decentralized \\ Allowlist}} & \multicolumn{1}{c|}{\full} & \multicolumn{1}{c|}{\makecell{Statistical \\ Measure}} & \multicolumn{1}{c|}{P} & \multicolumn{1}{c|}{Voting} & \multicolumn{1}{c}{\makecell{S \\ B}} &  \multicolumn{1}{c}{}\\

	\hline

	\end{tabular}
	\captionsetup[tabular]{singlelinecheck=off}
	\caption{A classification of the existing oracle implementations using the modular framework described in Section~\ref{overview_workflow}. \\ {\full} indicates the properties (columns) are implemented in the corresponding oracle (rows), and $\times$ indicates the property is not applicable. \\ $\dagger$ First Come First Serve $\star$\textit{The Market Creator assigns the designated reporter} $\star$$\star$  \textit{{The series of reported prices will be sent to requester without aggregation (See~\ref{provideranddatavetting})}}
	\label{tab:classification}}

\end{table*}

In Table~\ref{tab:classification}, we present a classification of several oracle implementations using the modular framework described in this section. This table showcases a wide variety of approaches, as well as some specialization on specific modules (\eg TownCrier and Deco on data source and Kleros on dispute resolution). We caution that blockchain projects can change how they work very quickly, new projects will emerge, and current projects will be abandoned. Table~\ref{tab:classification} has a limited shelf-life of usefulness, however the workflow itself (modules, sub-modules, and design choices) is based on general principles and intended to have long-lasting usefulness.



\section{Interacting with the blockchain}

While the initial inputs to an oracle are generally \textit{off-chain} (with the exception of pulling data from another smart contract) and the final output is by definition \textit{on-chain}, the oracle designer will choose to implement the intermediary modules---data feeder selection, aggregation and dispute resolution---as either off-chain or on-chain. Generally, on-chain modules are preferred for transparency and immutability, while off-chain modules are preferred for lower costs and greater scalability. 

To illustrate, Chainlink and NEST Protocols were ranked \#5 and \#7 respectively in gas usage among all DApps on Ethereum.\footnote{Based on Huobi DeFiLabs Insight on September 2020 ~\cite{huobiDeFiLabs}} This ranking was achieved mainly because they implement all modules fully on-chain. Later, Chainlink implemented an off-chain reporting (OCR) protocol~\cite{chainlinkocr} with the goal of reducing the gas costs associated with on-chain transactions. This protocol uses digital signatures to authenticate feeders and a standard (\eg Byzantine fault tolerant~\cite{castro2002practical}) consensus protocol between Chainlink nodes.

At some point, an oracle system must move on-chain and start interacting with the underlying blockchain. We assume for the purpose of illustration that Ethereum is the blockchain being used. Data flow from an off-chain module to a smart contract involves the following three components which we detail in this section. 

\begin{description}
	
	\item[\ref{offChainInfrastructure}] \textbf{Off-chain Infrastructure}: Assuming at least one module is off-chain, an infrastructure is required to monitor requests for oracle data from the blockchain, gather the data from the data sources, implement a communication network between data feeders, and create a final transaction to be sent to the blockchain infrastructure.

	\item[\ref{blockchainInfrastructure}] \textbf{Blockchain Infrastructure}: Off-chain infrastructure will pass the data as a transaction to blockchain nodes, which relay transactions and use a consensus algorithm agree on new blocks. The nodes run by miners are discussed in particular as they dictate the order of transactions in every block they mine.

	\item[\ref{smart_contract}] \textbf{Smart Contracts}: The transaction triggers a state change in a smart contract on the blockchain, typically a contract owned by the oracle which is accessible from all other contracts. Alternatively, the oracle could write directly into a data consumer's contract (called a \textit{callback}).
	
\end{description}

\subsection{Off-chain Infrastructure}  \label{offChainInfrastructure}

Depending on the oracle design, there can be different types of off-chain infrastructure. If financial data is pulled from Uniswap's oracle~\cite{uniswaporacle}, there is no off-chain infrastructure needed because the oracle is already a fully on-chain oracle. For other applications, off-chain infrastructure could consist of a single server (\eg TownCrier~\cite{zhang2016town}) or many nodes that intercommunicate through their own consensus protocol (\eg Chainlink OCR~\cite{chainlinkocr}). Availability and DOS-resistance~\cite{sonar2014survey} are core requirements of off-chain infrastructure, specially in oracle systems working with time-sensitive data and high update frequency. In this section we describe different possible components of the off-chain infrastructure.

\subsubsection{Monitoring the Blockchain}

For oracles that are capable of returning a custom data request made on-chain (called \textit{request-response oracles}), every data feeder needs to monitor the oracle's smart contract for data requests. The common implementation consists of a server subscribing to a blockchain node for specific events. 


\subsubsection{Connection to Data Source}

The data feeder requires a connection to the data source~\ref{data_sources} to fetch the desired data. This connection can be an entry point for an adversary to manipulate the data however it is possible to mitigate this issue by integrating message authentication  (recall source authentication in Section~\ref{source_authentication}). Examples include relaying HTTPS data (\eg Provable~\cite{bernanioraclize} via TLS\-Notary~\cite{tlsnotary}) or from trusted hardware enclaves (\eg TownCrier~\cite{zhang2016town} via Intel SGX~\cite{costan2016intel}). Vulnerabilities with the web-server or SGX itself~\cite{brasser2017software} are still possible attack vectors.

\subsubsection{Data Feeders Network}

In order to increase the scalability of the oracle network, multiple data feeders might aggregate their data off-chain (\eg Chainlink OCR~\cite{chainlinkocr}). In OCR, a leader is chosen from the participants to gather signed data points from other nodes. Once consensus is achieved on the aggregated set of data, the finalized data, accompanied by the signatures, is  transmitted to the blockchain node. This reduces the costs as only one transaction is sent to the blockchain, while maintaining similar security as having each chainlink nodes send the data themselves. 

Like any network system, availability is essential to the operation of the oracle. To illustrate, in December 2020, MakerDAO's oracle V2 had an outage due to a bug in their peer-to-peer data feeder network stack~\cite{makerdaooutage}. We do not summarize all the literature on peer-to-peer network attacks, but denial-of-service attacks~\cite{wood2002denial} and sybil-attacks~\cite{douceur2002sybil} are critical to mitigate to ensure the availability of the network and the oracle.

\subsubsection{Transaction Creation}\label{transaction_creation}

In order to submit data to a blockchain, the data feeder is required to construct a valid blockchain transaction that includes the requested data. This transaction must be signed with the data feeder's private key to be validated and authenticated on-chain. The data feeders must protect the signing keys from theft and loss~\cite{EBSC15}, as this key can be used to impersonate the oracle. 

Transactions compete for inclusion in the next block by offering different levels of transaction fees, known as the gas fee in Ethereum. In time-sensitive oracle applications, the relay must specify an appropriate amount of gas according to market conditions. For instance, on `Black Thursday' in March 2020~\cite{blackthursdayMaker}, the Ethereum network was congested by high fee transactions and some oracles failed to adjust their price feed. To mitigate this problem, the module which is responsible for creating the final transaction must have a \textit{dynamic gas} mechanism for situations where gas prices are rapidly climbing. In this case, pending transactions must be canceled, and new ones must be generated with higher gas price, which may take a few iterations to get in. Dynamic fees depend directly on the network state and require a connection to the blockchain node to estimate the adequate gas price.

In addition, the data feeder's sending address on the blockchain must have sufficient funds to be able to pay the estimated gas price. It is crucial for the availability of the oracle that the data feeders monitor their account balance as spam attacks might drain their reserves with high gas fees, as happened to nine Chainlink operators in September 2020~\cite{chainlinkdosattack}.

\subsection{Blockchain Infrastructure}  \label{blockchainInfrastructure}

In this section, we discuss the blockchain infrastructure that is required by any entity interacting with the blockchain. While this infrastructure is not specific to oracles, we illustrate key points that can impact oracle availability. 

\subsubsection{Blockchain Node} 

A blockchain node relays transactions to the other nodes in the network for inclusion in the blockchain. The node is responsible for storing, verifying, and syncing blockchain data. The availability of nodes is very important for the oracle system, as a blocked node cannot send transactions. Extensive research on network partitioning attacks apply to decentralized networks, with the main objective of surrounding an honest nodes with the malicious nodes~\cite{vasek2014empirical,neudecker2015simulation,zhang2019double,heilman2015eclipse,henningsen2019eclipsing}. This results in the node believing it is connected to the blockchain network when it is not.

\subsubsection{Block Creation}\label{block_creation}
Transactions that have been circulated to the blockchain network are stored in each node's \texttt{mempool}. Mining nodes select transactions from their \texttt{mempool} according to their priorities (\eg by highest gas price as in Geth~\cite{geth}, while respecting nonces). Front-running attacks~\cite{eskandari2019sok,daian2020flash} try to manipulate how miners sequence transactions. For example, someone might observe an unconfirmed oracle transaction in the mempool, craft a transaction that profits from knowing what the oracle data will be, and attempt to have this transaction confirmed before the oracle transaction itself (called an \textit{insertion attack}~\cite{eskandari2019sok}). This might be conducted by the miner themself. In this case, it is called \textit{transaction reordering}, and the profit miners stand to make from doing this is termed \textit{Miner Extractable Value (MEV)}~\cite{daian2020flash}). Other nodes or users on the network who can act quickly and offer high fees can also conduct front-running attacks. Users might also attempt a \textit{bulk displacement attack}~\cite{eskandari2019sok} that fills the consecutive blocks completely to delay reported data from oracles. There could be a profit motive for this attack if the oracle data becomes expired, or if the data feeder's collateral is slashed and redistributed to the attacker. 

Research on MEV (\eg Flashbots~\cite{flashbots}) has shown the possibility of new type of attacks based on reordering the transactions, such that if there's a high profit for changing the order of some transactions in a (few) blocks, miner is incentivize to use his hash rate to perform a reorganization attack\footnote{Also referred to as \textit{Time-bandit attacks}~\cite{daian2020flash}}~\cite{lin2017survey}, and profit from the execution of the newly ordered transactions. For instance, Uniswap uses the last price in a block to determine the average price (TWAP), in which a miner can add new trades while reordering the past trades with the goal of manipulating the price average to profit on other applications that uses Uniswap as price oracle. 

\subsubsection{Consensus} \label{consensus}

The goal of the consensus algorithm used in the blockchain is to verify and append the next block of transactions to the blockchain. If the nodes do not come to agreement on a state change, a fork in the network happens with different nodes trying to finalize different forks of the blockchain. Given the network is decentralized, short-lived forks happens frequently in the network that generally are resolved within a few blocks~\cite{neudecker2019short}. All valid transactions in the abandoned fork will eventually be mined in the main chain, likely in a new order (called \textit{reorganization} or a \textit{reorg}). 

A reorg opens the possibility of attacks by using known, unconfirmed, transactions from the abandoned fork. To illustrate, consider Etheroll~\cite{etheroll}, an on-chain gambling game where users bet by sending a number that payouts if it is smaller than a random number determined by an oracle. To prevent front-running from the mempool, the Etheroll oracle would only respond when a bet was in a block. Despite this mitigation, in April 2020, the Etheroll team detected an ongoing front-running attack on their platform~\cite{etherollincident}. The attacker was betting rigorously and waiting for small forks---collected by Ethereum in \textit{uncle blocks}---where the original bet and oracle's random number response were temporarily discarded by the reorg. The attacker would place a winning bet with a high fee to front-run the original bet and eventual inclusion of the oracle's transaction in the reorganized chain. A general principle of this attack is that even if oracle data bypasses the mempool and is incorporated directly by miners, front-running through reorgs is still possible. 

There are two solutions to front-running through reorgs. The first is to delay the settlement of the bet by a few blocks to prevent issues caused by small reorganization forks. The second is to incorporate a hash of the request (\eg \textit{request-id}) in the response to prevents the request (\eg bet) from being swapped out once the response (\eg random number) is known.

Other consensus attacks~\cite{gramoli2020blockchain,bissias2016analysis,henningsen2019eclipsing} exist but are less related to oracles. We omit discussion of them.

\subsection{Smart Contracts}  \label{smart_contract}

Although oracles are usually designed to be the source of truth for on-chain smart contracts, some smart contracts can also be used as oracles by others even though they were not designed with the oracle use-case in mind. To expand this idea, oracles could be a \emph{'an end in itself'}, which is to say they are designed specifically to be used as a source of truth. These oracles fetch the data from external sources(\ref{data_sources}) and make it available on-chain (\eg PriceGeth~\cite{eskandari2017feasibility}).

By contrast, \emph{a means to an end} oracle is a contract that produces useful data as a byproduct of what it is otherwise doing. Examples are on-chain markets and exchanges like Uniswap and other automated market makers (AMMs). The markets are designed for facilitating trades but provide pricing information (\emph{price discovery}) that can be used by other contracts (\eg margin trading platforms) as their source of truth.

In this section we dive deeper in the relationship between the oracle's smart contract and the data consumer smart contract. We start by defining possible interaction models, and then discuss specific issues related to the oracle's contract and the consumer's contract. 

\subsubsection{Oracle Interaction Models}

A distinction in the oracle design is whether the interaction between with the consumer's contract is implemented as a \textit{feed}, a \textit{request-response}, or the related \textit{subscribe-response}.

A \textit{Feed} is a smart contract system that publishes the data for others to use. It does not require any requests to fetch the data and using an interval to update the data on its smart contract (\eg Maker DAO Oracle~\cite{MAKERDAOOracle}). From a technical aspect, in order to use a feed oracle, the data consumer smart contract only needs to query the oracle's smart contract and no additional transactions are needed. 

The \textit{Request-Response} model is similar to a client-server API request on traditional web development. The requester must send a request to the oracle's smart contract, which then is picked up by the off-chain module of the oracle to fetch the requested data from the data source. The data is then encapsulated in a transaction and sent back to the data requester smart contract through the oracle's smart contract. Due to the nature of this design, at least two transactions are needed to complete the work flow, one from the requester and another for the responder.

The \textit{Subscribe-Response} model is similar to Request-Response with one main difference, the request does not need to be in a transaction. If there is pre-arranged agreement, the oracle will watch for emitted events from the requester smart contract and respond to the requests. Alternatively, the requester is allowed to read the feed through an off-chain agreement (\eg API3~\cite{benligiraydecentralized}).

\subsubsection{Oracle's Smart Contract}\label{oracle_smart_contract} 

In the oracle designs that implement some of the modules on-chain, the oracle's smart contract could include data feeder selection (Section~\ref{data_feeder_selection}), aggregation (Section~\ref{aggregation}), and dispute resolution (Section~\ref{dispute_phase}). In addition to these modules, the oracle's smart contract can be used as the data feed storage for other smart contracts to read from, or to authenticate the oracle's response on the consumer smart contract.  In the \textit{feed} model, the oracle's smart contract is where the consumer fetches the oracle data from. In the \textit{Request-Response} model, the data consumer smart contract (defined below in Section~\ref{data_consumer}) requires knowledge of the oracle's smart contract's address in advance, for the initial request and also verification of the oracle's response. For the rest of this section, we discussion potential attacks on the oracle's smart contract.


\paragraph{Implementation Flaws}
There are many known smart contract vulnerabilities that have been extensively discussed~\cite{attacksonethereum,chen2020survey} and possibly could affect the legitimacy of the oracle system. 

In many DeFi projects, a common design pattern is to use on-chain markets, such as Uniswap, for the price oracle, however, these systems were not designed to be used as oracles and are prone to market manipulation. The end result is that currently, the most prevalent attack vector in DeFi is oracle manipulation~\cite{defiattacksreport}. To illustrate this attack, consider the lending (and margin trading) platform bZx. It fetched prices from KyberSwap, a decentralized exchange, to calculate the amount of collateral of one cryptoasset is needed to back the loan of a different asset. In one attack on bZx~\cite{bzxPeckShield}, the attacker used a \textit{flash loan} to manipulate KyberSwap's sUSD/ETH exchange rate. The attacker then borrowed ETH with insufficient collateral because the bZx contract believed the collateralized sUSD was worth much more than it actually was. When the attacker absconded with the borrowed ETH, forgoing its collateral, and then unwound its other positions and repaid the flash loan, it profited at bZx's expense. Arguably bZx (the data consumer) is the flawed contract but the ease in which KyberSwap (the oracle contract) could be manipulated was not well understood at the time either. In reaction, decentralized exchanges embraced their role as a price oracle and hardened themselves against price manipulation by using aggregation methods like the \textit{Time-Weighted Average Price} (TWAP) (described in Section~\ref{aggregation}). 

\paragraph{Governance} 
In order to remove the centralization of control in many DeFi projects, a governance model is introduced that uses a native token for voting and staking. The governance model for an oracle could propose, vote, and finalize changes to system variables like the approved data feeders on the oracle's allowlist or various fees. 

While a decentralized governance model removes the trust in a central entity, it does not remove the possibility of a wealthy entity (a \emph{whale}) taking control of the system by accumulating (or borrowing~\cite{qin2020attacking}) enough tokens to pass their proposals. In addition, logical issues in the governance implementation could result in tricking the voters into approving a proposal that has malicious consequences~\cite{nexusmutualbug}. 

As an example, in the MakerDAO platform, MKR token holders can vote to change parameters related to Maker's oracle module~\cite{MAKERDAOOracle}. An attacker in October 2020, used a flash loan to borrow enough MKR tokens to pass a governance proposal, aimed to change the list of consumer smart contracts and obtain read access to the Maker's oracle~\cite{makerdaoflashloanattack}. It could be more dangerous if the attacker planed to change the other parameters of the oracle such as \textit{Whitelisted data feeders} or \textit{bar} parameter : the sufficient number of data feeders for data feeder selection module. Potentially an attacker may pay a bribe to the MKR holders to buy their votes, or use a \emph{Decentralized Autonomous Organization (DAO)} to pay for the votes without having ownership of the tokens~\cite{darkdao}.

\subsubsection{Data Consumer Smart Contract} \label{data_consumer}


The final point in the oracle workflow is the smart contract that needs the data for its business logic. Aside from any possible code vulnerabilities in this smart contract, there are common implementation patterns concerning the oracle workflow. 

In the \textit{feed} model, the data consumer smart contract relies on oracles to fetch the required data in order to function as intended. It is essential to use oracles with multiple data feeders and a proper aggregation methods. To illustrate the importance, consider the lending service Compound~\cite{compoundPriceFeed} which initially only used Coinbase Pro as their data feeder without any aggregation mechanisms~\cite{compoundcoinbasepro}. In November 2020, a faulty price feed on Coinbase Pro, resulted in undercollarization of Compound loans and a liquidation of \$89 million dollars of the collateral. This could have been prevented by using an oracle with sufficient data feeders and a proper aggregation mechanism. 

Due to the commonality of this issue, there has been some Ethereum Improvement Proposals (EIPs) to standardize the interface of the oracles implementing a \textit{feed} (\eg EIP-2362~\cite{eip2362}). An interface would allow data consumer smart contracts to easily switch between feeds or use multiple oracle feeds in their logic. 

In the \textit{request-response} model, the data consumer smart contract sends a request for specific data to the oracle's smart contract. In some projects this request contains more information like the data feeder selection method, aggregation algorithm and parameters for dispute phase (\eg Service Level Agreement in Chainlink). It is crucial that the data consumer smart contract, verifies the authenticity of the oracle response. Failure to verify the oracle's response could result in malicious data injection in the data consumer smart contract. To illustrate, the insurance service Nexus Mutual~\cite{nexusmutualbug} implemented an oracle's response function (or callback) without any proper access control. This opened the possibility of unauthorized entities providing data updates which would be wrongfully assumed to have originated from the oracle's smart contract.



\section{Concluding Remarks}
In this paper, we described a specialized modular framework to analyze oracles. After our systematization, we present the following discussion points and lessons learned from our work. 

\begin{enumerate}
    \item Many oracles projects introduce their own governance tokens that are used to secure the oracle system (\eg through staking). Two conditions seem necessary: the market capitalization of the token stays material and the token is evenly distributed. More consideration should be given to leveraging an existing token with these properties (even a non-oracle token) instead of creating new specialized tokens~\cite{vitalikuni}. Also a collapse in the value of the governance token threatens the entire system.

    \item Oracle systems with on-chain modules are expensive to run on public blockchains like Ethereum, which prices out certain use-cases that consume a lot of oracle data but do not generate proportional amount of revenue (\eg Weather data). 


    \item Diversity in software promotes resilience in the system. If the oracle market coalesces  behind a single project, a failure within this project could cause cascading failures across DeFi and other blockchain applications.  

    \item While determining the profit from corrupting the oracle is a promising approach to thwarting manipulation (by ensuring the cost of corruption is greater), one can never capture the full extent of the potential profit. Attackers can profit outside of Ethereum by attacking oracles on Ethereum~\cite{FoBo19}.

\end{enumerate}

In summary of this paper, the framework we present facilitates a modular approach in evaluating the security of any oracle design and its associated components that exist today or to be implemented in the future. As an example, the level of centralization can be measured using choke points such as aggregation~\ref{aggregation}, or how the data is proceeded to the blockchain~\ref{offChainInfrastructure}. In order to design a secure oracle, all modules must be rigorously stress tested to make sure it cannot be gamed by participants or malicious actors. In addition, many security auditors and analysis tools are specialized in detecting oracle-related attacks through code review of the smart contracts. Specially with the rise of DeFi smart contracts, the importance of a secure oracle system remain a paramount component of the decentralized blockchain ecosystem.


\begin{acks}

J. Clark acknowledges support for this research project from (i) \grantsponsor{chaire}{The Chaire Fintech: AMF -- Finance Montréal}{https://chairefintech.uqam.ca}, (ii) \grantsponsor{nserc}{National Sciences and Engineering Research Council (NSERC)}{https://www.nserc-crsng.gc.ca} through the NSERC, Raymond Chabot Grant Thornton, and Catallaxy Industrial Research Chair in Blockchain Technologies \grantnum[https://www.nserc-crsng.gc.ca/Chairholders-TitulairesDeChaire/Chairholder-Titulaire_eng.asp?pid=1045]{nserc}{}, and (iii) an NSERC Discovery Grant \grantnum{nserc}{}.

\end{acks}


\bibliography{bib/bib}



\end{document}